\documentclass[aps,prl,reprint,groupedaddress]{revtex4-1}

\usepackage{graphicx}
\usepackage{color}
\usepackage{subcaption}
\usepackage{mathtools}
\usepackage{cleveref}

\newcommand{\vb}[1]{\mathbf{#1}}
\newcommand{\tens}[1]{\underline{\underline{\mathbf{#1}}}}

\begin{document}

\title{Nernst Effect in Magnetized Hohlraums}

\author{A.S. Joglekar}
\email{archisj@umich.edu}
\affiliation{Department of Nuclear Engineering and Radiological Sciences,  University of Michigan, Ann Arbor, MI, USA}
\author{C. P. Ridgers}
\affiliation{York Plasma Institute, University of York, UK}
\author{R. J. Kingham}
\affiliation{Blackett Laboratory, Imperial College, London, UK}
\author{A.G.R. Thomas}
\email{agrt@umich.edu}
\affiliation{Department of Nuclear Engineering and Radiological Sciences,  University of Michigan, Ann Arbor, MI, USA}
\date{\today}

\begin{abstract}
We present  nanosecond timescale Vlasov-Fokker-Planck-Maxwell modeling of magnetized plasma transport and dynamics in  a hohlraum with an applied external magnetic field, under conditions similar to recent experiments. Self-consistent modeling of the kinetic electron momentum equation allows for a complete treatment of the heat flow equation and Ohm's Law, including Nernst advection of magnetic fields. In addition to showing the prevalence of non-local behavior, we demonstrate that effects such as anomalous heat flow are induced by inverse bremsstrahlung heating. We show magnetic field amplification up to a factor of 3 from Nernst compression into the hohlraum wall. The magnetic field is also expelled towards the hohlraum axis due to Nernst advection faster than frozen-in-flux would suggest. Non-locality contributes to the heat flow towards the hohlraum axis and results in an augmented Nernst advection mechanism that is included self-consistently through kinetic modeling. 
\end{abstract} 

\maketitle

Indirect drive inertial confinement fusion (ICF) is accomplished through the compression of a fuel capsule due to the exposure of an ablator material to radiation. This radiation is a result of laser heating of a hollow-cylinder made of a high Z material. At the National Ignition Facility (NIF), a gold hohlraum is irradiated in a precise arrangement of 351 nm laser beam cones. The walls of the hohlraum are heated to an electron temperature of approximately 3-5 keV with the goal of emitting blackbody-like radiation of 300-500 eV. The resulting x-ray bath is intended for uniform compression of the fuel capsule to fusion conditions.  \cite{Lindl2004}

There has been recent interest in the role of applied magnetic fields in high-energy-density plasmas  \cite{Chang2011,Hohenberger2012,Fiksel2014} for inertial fusion energy applications. The Magneto-Inertial Fusion Electric Discharge System has been developed in order to provide steady state magnetic fields for long time-scales relative to the experiments. A recent experiment on the Omega Laser Facility with a 7.5~T external axial magnetic field imposed on an Omega-scale hohlraum measured a rise in observed temperature along the hohlraum axis \cite{Montgomery2015}. Recent modeling showed that hot electrons from laser-plasma interactions \cite{Regan2010} can be guided through the hohlraum, rather than the capsule, using such fields \cite{Strozzi2015}.

From a complete treatment of Ohm's Law, it has previously been shown that electron heat transport can advect such magnetic fields through the Nernst effect \cite{Nishiguchi1984,Haines1986,RidgersPRL2008,Willingale2010,Joglekar2014,Li2013} in addition to well-known MHD processes like ``frozen-in-flow'' and resistive diffusion. Dimensionless numbers that compare the ratio of the magnitudes of the Nernst  term in Ohm's law to  that due to  bulk plasma flow, $R_N \gg1$ \cite{RidgersPRL2008}, and the Hall term, $H_N \gg1$ \cite{Joglekar2014},  suggest that Nernst convection should be the dominant mechanism for magnetic field transport in a hohlraum. Such a hot and semi-collisional environment is, however, also rich in non-equilibrium effects that may complicate the magnetic field dynamics.

Laser heating of the plasma results in steep temperature gradients, typically $\mathcal{O}(3~ \text{keV} / 50~ \mu\text{m})$. The collisional mean-free-path of a 3 keV electron is $ \mathcal{O}(10 ~ \mu$m), depending on the plasma density. Since $\lambda_\text{mfp}/L < 100$, non-local effects can be expected to be important \cite{Gray1977}. The steep temperature gradients that occur due to the intense laser heating in a hohlraum have been shown to result in non-local heat flow \cite{Gregori2004,Hawreliak2004}. Careful consideration of the population of electrons with $2v_{\text{th}}<v < 4v_{\text{th}} $ is required as these carry most of the heat. Additionally,  inverse-bremsstrahlung heating of a plasma has been shown \cite{Langdon1980,Liu1994} to not only lead to deviations from classical transport, prescribed by Braginskii's transport equations \cite{Braginskii1965}, but also new transport terms \cite{Ridgers2008}. Both non-local transport and laser heating result in modifications to the shape of the distribution function and therefore non-equilibrium behavior, which mean that classical transport approximations break down. In order to avoid classical transport approximations, a \textit{kinetic approach} is necessary. Kinetic modeling allows for the modeling of magnetic field dynamics through a self-consistent and generalized Ohm's Law that can be derived without distribution function approximations. 

In this Letter, we demonstrate  Vlasov-Fokker-Planck-Maxwell (VFPM) simulations of a 
magnetized, full hohlraum-scale plasma including ray-tracing of an Omega-like laser configuration over a nanosecond time-scale. The hohlraum is considered without an ICF capsule but a gas fill throughout. Note that radiation transport and laser-plasma interactions are neglected in these calculations, which may change our results if they were included. With the use of \textsc{IMPACTA} \cite{Kingham2004,Thomas2009}, we studied the effect of non-equilibrium electron kinetics on thermal energetic and magnetic field dynamics of a Omega-scale hohlraum with an externally imposed 7.5~T magnetic field. We found that significant proportions of the total heat flow are non-local. Additionally, the presence of inverse bremsstrahlung heating resulted in anomalous heat flow towards the over-dense plasma of the hohlraum wall. Therefore, the diffusive heat flow from the laser-heated regions is not an adequate description of the thermal energy dynamics. The heat flows from the laser heating move the externally imposed magnetic field through Nernst advection. To examine the effects of Nernst advection in relation to the plasma bulk flow, we show modeling without an electron contribution to the transport of  magnetic field in Ohm's Law for comparison. 

We find that magnetic field transport due to Nernst flow results in significantly faster field cavitation than that is possible via frozen-in-flux. Magnetic field cavitation occurs due to heat flow down the density and temperature gradient, which is shown to be non-local. Retention of the distribution function allows for accurate modeling of the magnetic field cavitation because the local approximation to the Nernst velocity underestimates the true convection velocity by a factor of 2. Nernst flow into the over-dense region causes magnetic flux pile-up at the walls and results in magnetic field amplification by a factor of 3. Magnetic flux pile-up does not occur with only plasma bulk flow present as there is a negligible amount of plasma bulk flow toward the wall from the laser heated region.

The Vlasov-Fokker-Planck (VFP) equation for electrons is solved coupled with Ampere's and Faraday's Laws and a hydrodynamic ion fluid model to describe the plasma. The code we use, \textsc{Impacta} \cite{Kingham2004,Thomas2009}  uses a Cartesian tensor expansion, with the distribution function expanded as $f(t,\vb{r},\vb{v})=f_0 + \vb{f}_1\cdot\hat{\vb{v}} + \tens{f}_2:\hat{\vb{v}} \hat{\vb{v}} +\dots$, where $\hat{\bf v}(\theta,\phi)$ is a unit velocity vector. This expansion can be truncated in a collisional plasma, as collisions tend to smooth out angular variations in the  distribution function, resulting in a close to isotropic distribution, represented by $f_0$. Higher orders are successively smaller perturbations, $\tens{f}_2 \ll \vb{f}_1 \ll f_0$ etc.  In the classical limit that $f_0$ is a Maxwell-Boltzmann velocity distribution, \textsc{Impacta} has been shown to agree with Braginskii's transport equations \cite{Kingham2004}. In \textsc{Impacta}, terms up to and including $\tens{f}_2$ can be retained. These simulations, however, are collisional enough such that $\tens{f}_2$ may be neglected to an error $\mathcal O (\lambda_\text{mfp}/L)^2$.

\begin{figure}[h]
	\centering
	\includegraphics[width=\columnwidth]{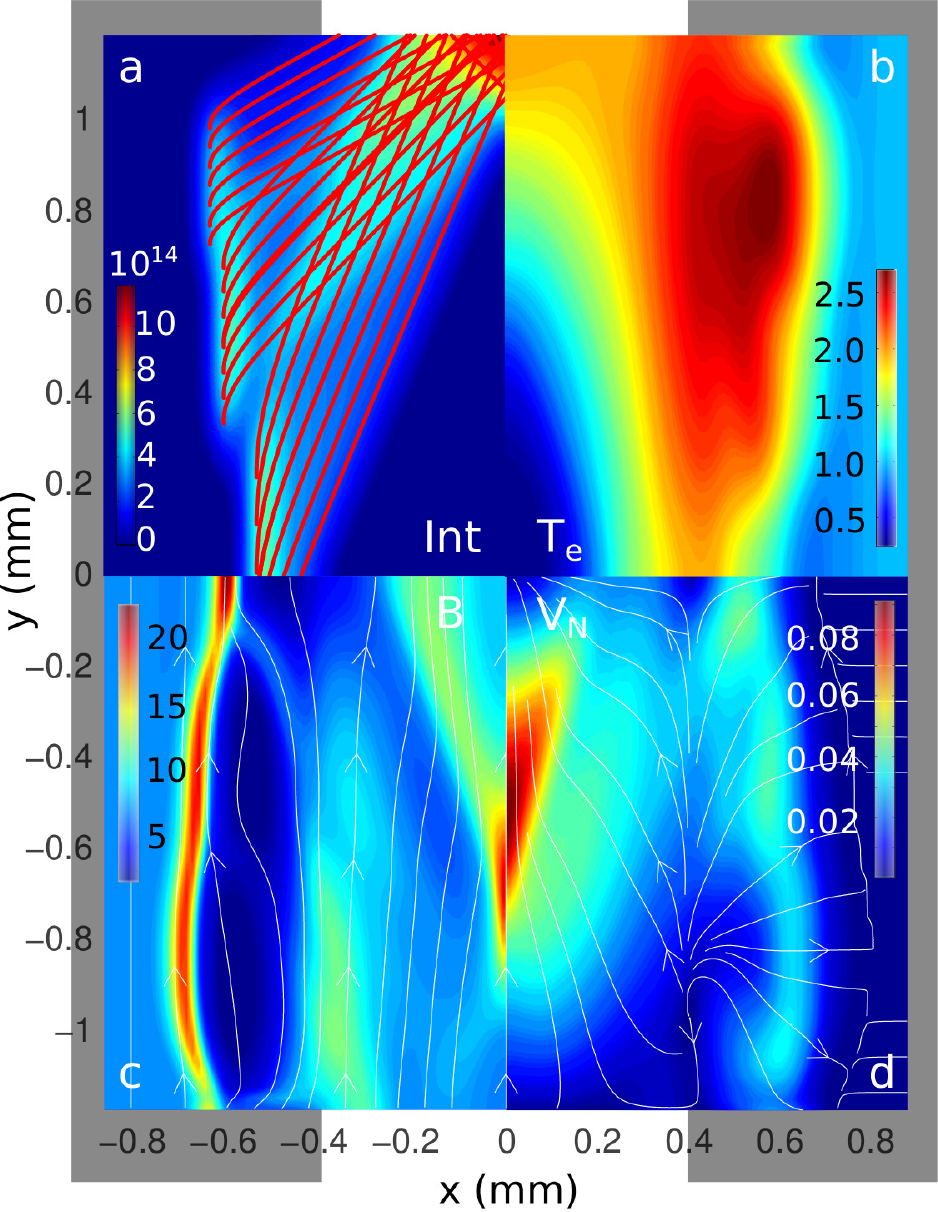}
	\caption{(a) Ray tracing profile overlaid onto laser intensity profile (W/cm$^2$) at t = 0. (b) Electron Plasma Temperature (keV), (c) Externally applied magnetic field (T), (d) Nernst Velocity ($v_N/v_\text{th0}$) at t = 250 ps.} 
	\label{hohlraum}
\end{figure}

A 2-dimensional slice of a hohlraum is modeled in the x-y plane where the y-axis represents the longitudinal axis of the hohlraum and the fuel pellet would sit at the origin. The hohlraum walls are represented by a dense, high-$Z$ plasma located at approximately $x = \pm 800 ~\mu$m, and the gas fill by a low-$Z$ plasma, with the overall $Z$ distribution described by the function $Z(x,y) = 59.25+19.75 \tanh(\frac{x-750}{40})$. Electron number density is described by the function $n_e(x,y) = (2.98+2.93 \tanh(\frac{x-750}{40}))\times10^{22}\;\text{cm}^{-3} $. The initial uniform temperature was $k_BT_{e0} = 160$ eV. The initial uniform magnetic field was $\mathbf{B}_0(\hat{y}) = 7.5$ T and $\ln\Lambda_{ei}=5.4$. To convert from the normalized units, $n_{e0} = 5 \times 10^{20}$ cm$^{-3}$ and $v_{\text{th}0}/c = 0.025$ are used. . The laser parameters are designed to resemble  those of ref. \cite{Montgomery2015}. The ray tracing package tracks the three beam cones that enter at 21, 42, and 59 $\deg$ from the axis, to their respective refraction points and allows for some reflection. 
  
The rays and the initial heating profile are shown in \cref{hohlraum}a. \Cref{hohlraum}b shows the temperature profile after 300~ps of laser heating. \Cref{hohlraum}c shows the cavitation and amplification in the in-plane magnetic field profile caused by intense laser heating.  The Nernst velocity, shown in \cref{hohlraum}d, is directed towards the hohlraum axis in the low density gas fill and into the hohlraum wall in the Au plasma. Throughout the rest of this Letter, we show that the Nernst flow is primarily responsible for the magnetic field profile seen in \cref{hohlraum}c.

It should also be noted that while self-generated magnetic fields, $B_z$ in this geometry, are also included in this study, their dynamics are not the subject of this Letter. Many of the effects discussed here, however, can be applied to the self-generated fields and will be described in detail separately.

\begin{figure}[h]
\centering
\includegraphics[width=\columnwidth]{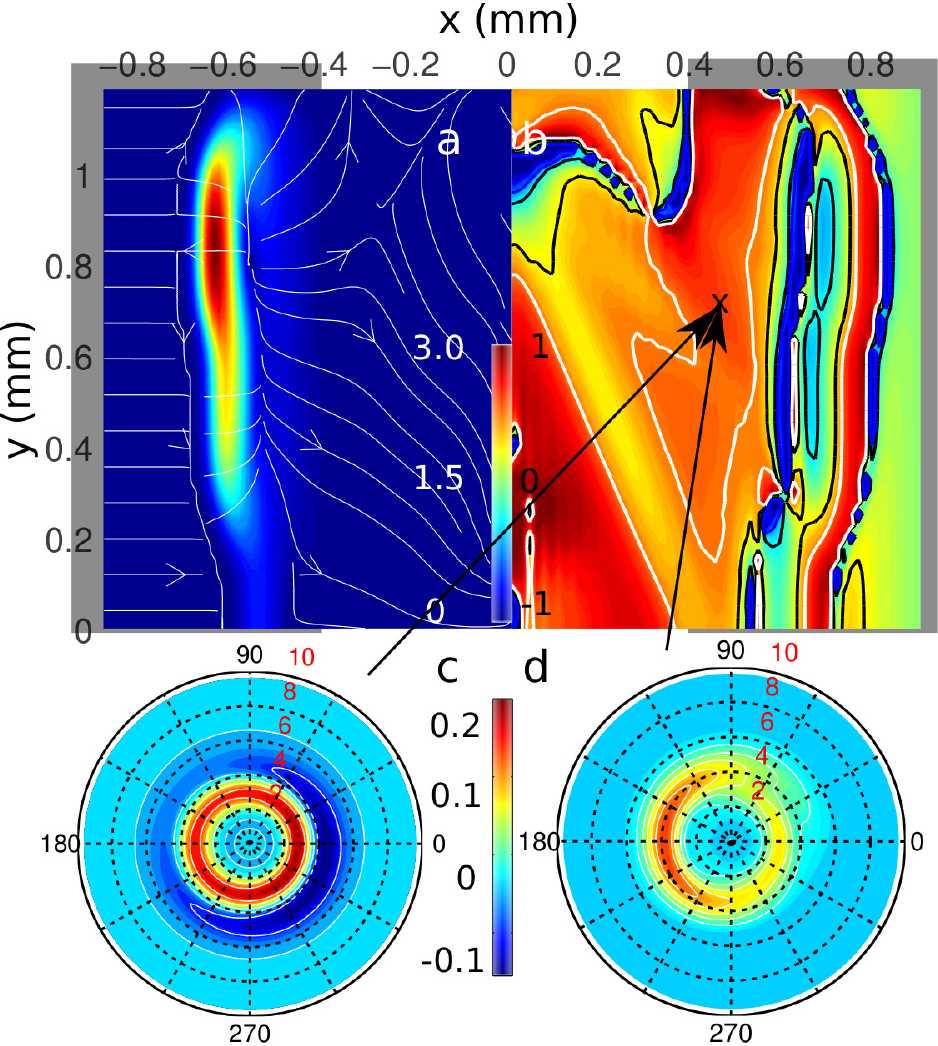}
\caption{(a) Heat flow ($m_e n_{e0} v_{\text{th}0}^3$) (b) $1-q_\text{\cref{clQ}}/q_\text{code}$ \\ (c) $v^5 (f_\text{code}-f_\text{MB})$ (d) $v^5 (f_\text{code}-f_\text{SG})$\\ $@~x = 0.4$ mm, $y = -0.6$ mm,\\ $m = 2.625$ , $t = 100$ ps.}
\label{fig::heat}
\end{figure}

Inverse bremsstrahlung heating of the plasma results in a super-Gaussian electron distribution \cite{Langdon1980}, which  consequently modifies the transport coefficients \cite{Ridgers2008,Bissell2013} and even introduces new terms including an anomalous heat flux up a density gradient $\mathbf{q}_n$, represented by the last term in \cref{clQ}; 
\begin{equation}
\mathbf{q}_e = - \frac{T_e}{e} \underline{\underline{\psi}}^\prime \cdot \mathbf{j}- \left(\underline{\underline{\kappa}}+n_e\underline{\underline{\phi}}\right) \cdot \nabla T_e  - T_e\underline{\underline{\phi}} \cdot \nabla n_e\;, \label{clQ}
\end{equation}
where $\psi$, $\phi$ and $\kappa$ are transport coefficients as described in reference \cite{Ridgers2008}. $\mathbf{q}_n$ increases as $m>2$ increases, where $m$ is the power of the super-Gaussian distribution function defined by $f_\text{SG}(v) = C(m) {n_e}/{v_\text{th}^3}\exp\left(-\left({v}/{\alpha_e v_\text{th}}\right)^m\right)$ where $\alpha_e = [3\Gamma(3/m)/2\Gamma(5/m)]^{1/2}$ and $C(m)=m/4\pi\alpha_e^3 \Gamma(3/m)$. 

In these simulations, by finding the best fit of a super-Gaussian  distribution to $f_0$, $m$ reaches a maximum of 3.1 near the centers of the laser heated regions, but  varies spatially and temporally, thus requiring the preservation of the distribution function at each point throughout the simulation for accurate calculation of the heat flow. Using the theory detailed in refs. \cite{Ridgers2008,Bissell2013}, the heat flow can be modified in a hydrodynamics code to include this effect. However, the  distribution is not precisely a super-Gaussian due to other effects such as non-locality and therefore this fix remains an approximation.

We examine the relative magnitudes of the real heat flow, and the classical heat flow calculated using all three terms that form the full post-processed heat flow from \cref{clQ} that includes anomalous heat flow. Calculation of the anomalous heat flow as a function of the best-fit distribution function, table look-up, and pressure gradient shows that there is heat flow into the hohlraum wall due to the $\phi \nabla P_e$ term and this approximately results in a 10\% correction to the diffusive heat flow i.e. $\kappa \nabla T_e$. 

A majority of the  disagreement between the heat flow from the code and the heat flow from the post-processed modified classical transport theory is due to the strongly non-local heat flow that is prevalent in the hohlraum. \Cref{fig::heat}b shows a 2D profile of a metric for quantifying the magnitude of the discrepancy between the two heat flows, described by the relative difference between the classical and  calculated  heat fluxes, $1-q_\text{\cref{clQ}}/q_\text{code}$. 

The regions within the black contours have $\pm 25\%$ agreement between the two heat flows. The white contours correspond to regions of high non-locality where the classical transport calculation is an underapproximation, while the blue contours correspond to regions where the heat flow is significantly overcalculated by classical transport. Heat flow from regions near the temperature hotspots, $\pm 50~\mu$m, is overestimated by the classical calculation while the heat flow further away from the hotspots, $\pm 200 ~\mu m$, is underestimated, as expected from the existence of non-locality. The regions of relative agreement are $\pm ~ 50-200 \mu$m from the hot spots. Due to the laser heating, the thermal electron mean-free-path increases,  $\lambda_\text{mfp}/L > 0.02$, suggesting that non-local heat flow becomes prevalent in the laser heated region.
 
Consideration of the in-plane electron distribution function $f(\theta,v) = f_0+f_{1x} \hat{v}_x+f_{1y} \hat{v}_y$ can show the significance of inverse-bremsstrahlung heating and non-locality. Since $\mathbf{q} \propto \int v^5\,f(\theta,v)\,\hat{\mathbf v}(\theta,\phi)\,dv \sin\theta d\theta d\phi$, the important contributions to the heat flow may be best illustrated by the function $v^5f(\theta,v)$. \Cref{fig::heat}c and d show the difference between the calculated distribution $v^5f$ and (c) a Maxwell-Boltzmann $v^5f_\text{MB}$ and (d) a super-Gaussian with best fit to $m$, both with $T_e$ equal to that of $f(x=0.4,y=-0.6)$.  \Cref{fig::heat}c shows that $f>f_\text{MB}$ in the region $2<v_\text{th}<4$ and $f<f_\text{MB}$ in the region  $4<v_\text{th}<6$, which is characteristic of inverse-bremsstrahlung heating. Calculating the heat flow contribution difference between the real distribution and the best-fit super-Gaussian ($m\approx2.625$ in this case), shows that the inverse-bremsstrahlung model does not replicate the distribution function fully due to anisotropy from the flow  and non-local effects. The enhanced tail and shifted center in the $180^\circ$ direction is characteristic of the (non-local) heat flow down the density gradient while the colder return flow is a result of the features in the $0^\circ$ direction.

As shown in ref. \cite{Haines1986}, the Nernst velocity is,
\begin{align}
\vb{v}_N &= \frac{\langle \vb{v} v^3 \rangle}{2 \langle v^3 \rangle} + \frac{\vb{j}}{e n_e} \label{eq:vNe}\\
&\approx \frac{\tens{\kappa}\cdot \nabla T_e}{5/2 P_e},\label{vnqe}
\end{align}

It can be shown for this geometry that $B_y$ has no field generation terms from the curl of Ohm's Law and therefore, can be transported through $(\vb{v}_N + \vb{C}) \times\vb{B}$ term in addition to resistive diffusion.
Over 0.5 ns, the simulation shows that there is magnetic field cavitation resulting in flux pile-up on the hohlraum axis and significant compression at the hohlraum wall due to the energy deposition from the laser. Pile-up of the magnetic flux results in a 25 T magnetic field, more than 3 times the strength of the initial 7.5 T field. 

\begin{figure}[h]
	\includegraphics[width=\columnwidth]{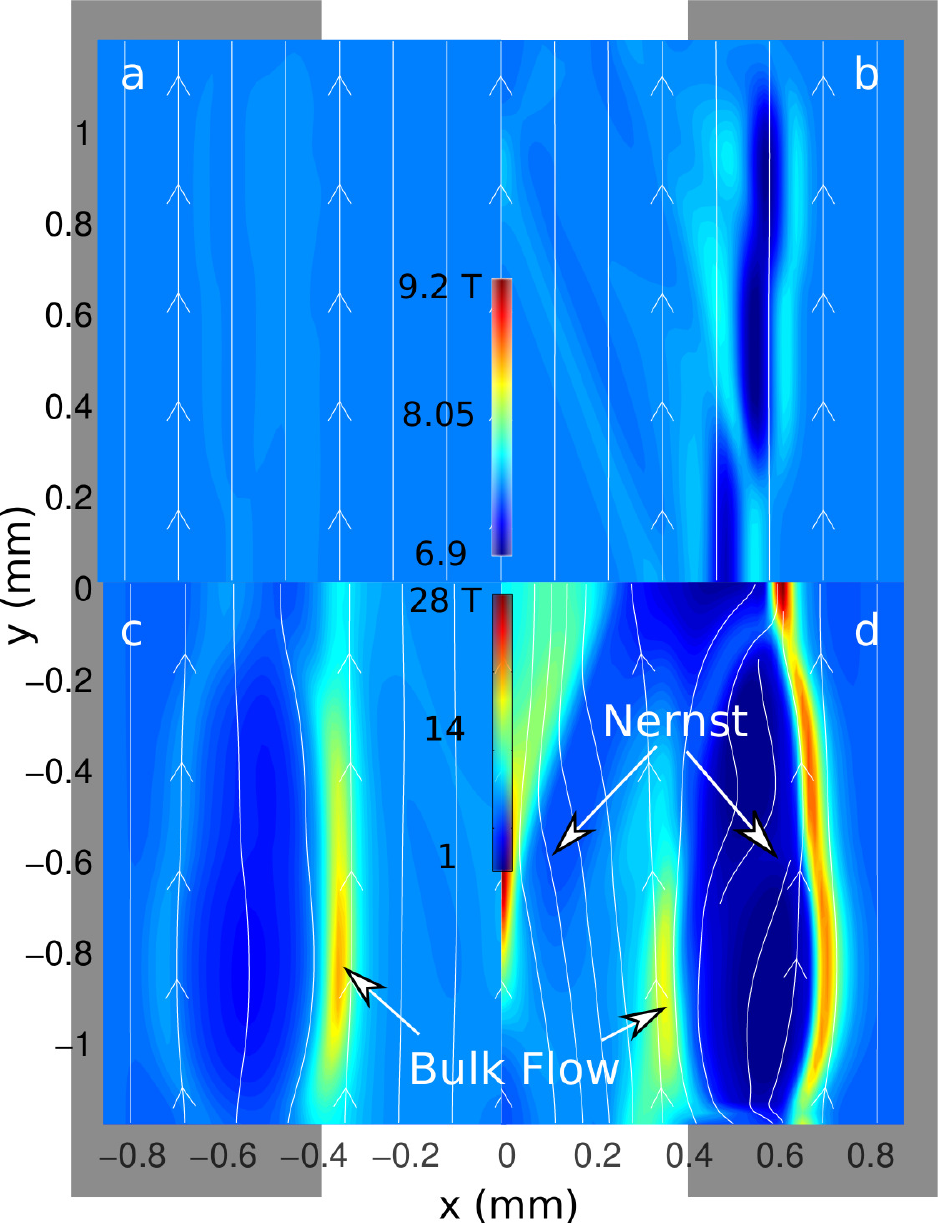}
	\caption{Magnetic field (T) after 50 ps with only plasma bulk flow (a) and full Ohm's Law (b). \\ Magnetic field after 400 ps with only plasma bulk flow (c) and full Ohm's Law (d)}
	\label{fields}
\end{figure}

In order to determine the effect of Nernst advection on the magnetic field evolution, simulations with and without the $ \vb{B} \times \vb{f}_1 $ term in the $\vb{f}_1$ equation were compared. This term is responsible for the interaction of kinetic electrons with the magnetic field. It is responsible for the Nernst and Hall terms in Ohm's Law as well as the Righi-Leduc effect in the heat flow equation. Simulations agree with the previous determination that $\vb j \ll \vb v_N$ because $H_N \ll 1$ and therefore, the Hall effect can be neglected. The magnetic field after 50 ps without and with full Ohm's Law treatment is shown in \cref{fields}a and \cref{fields}b, respectively.  The laser heated region results in magnetic field cavitation in both cases but the magnitudes differ. It is not evident in \cref{fields}a since the field is only modified by a few percent by the plasma bulk flow. Thermal energy transport results in a more noticeable change immediately over 50 ps. 

An estimate of the time-scale for the plasma bulk flow to transport frozen-in magnetic fields to the center of the hohlraum is given by, $ \frac{r_H}{C_s} \approx \frac{r_H}{\sqrt{k_B T_e/M_i}} \sim 2 ~ \text{ns}.$ \Cref{fields}d shows that including the Nernst effect results in magnetic field cavitation on a faster time-scale than can be expected due to field advection only through bulk plasma flow in \cref{fields}c. In the case of a 7.5 T initial field strength, the magnetic field on the axis grows to 30 T within 0.5 ns. \Cref{fields}d also shows that the magnetic flux pile-up in the hohlraum wall occurs due to the Nernst effect. The field increases to a strength of nearly 25 T towards the hohlraum wall.

\begin{figure}[h]
	\centering
	\includegraphics[width=\columnwidth]{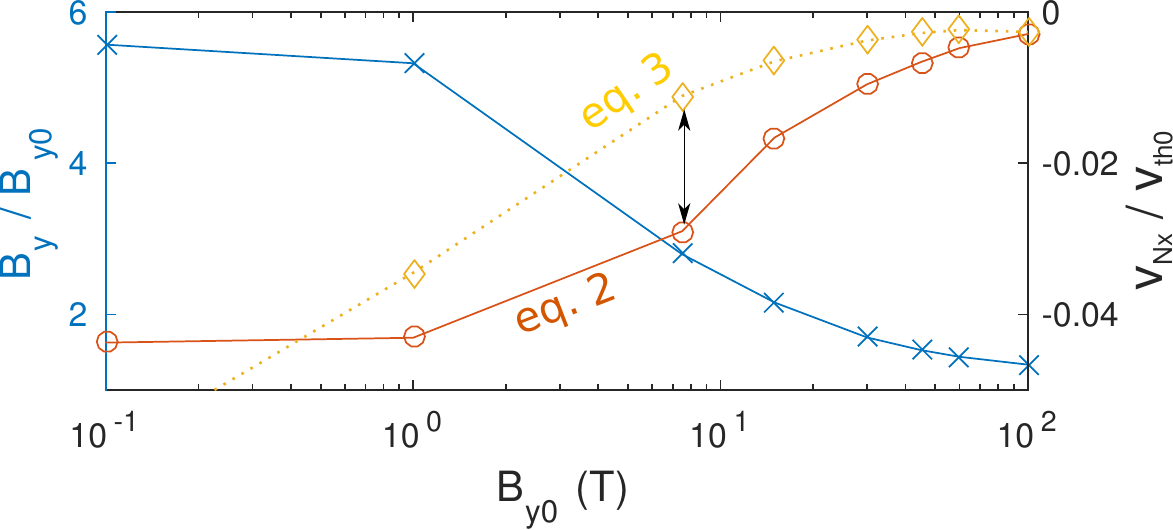}
	\caption{The magnetic field at the hohlraum axis decreases as applied field strength is increased because the Nernst effect is mitigated at higher magnetic field strengths, preventing the magnetic field from accumulating near the hohlraum axis. (t = 300 ps). The discrepancy between the approximated and exact Nernst velocity also decreases.}
	\label{fig::VNB}
\end{figure}

We also ran a series of simulations with varying initial applied  field $B_{y0}$ to understand how the field strength affects the hohlraum dynamics. \Cref{fig::VNB} shows the results of the magnetic field cavitation study for increasing field strengths suggesting that as 
\begin{equation}
\lim_{B_{y0}\rightarrow\infty} B_{y-\text{axis}}/B_{y0} = 1.
\end{equation}
The maximum value of $v_N$ in the domain of magnetic field advection towards the axis ($-0.5~ \text{mm} < x < 0.5 ~\text{mm}$) is chosen. This trend can be explained by the observed reduction in the Nernst velocity towards the hohlraum axis as the magnetization increases (also shown in \cref{fig::VNB}), which quenches magnetic field transport.  These $v_N(\omega\tau)$ curves are in line with other predictions \cite{Haines1986,Ridgers2008,Joglekar2014} that $v_N \propto 1/\omega\tau$ for $\omega\tau\gg1$.  \Cref{fig::VNB} also shows that the exact Nernst velocity from \cref{eq:vNe} is consistently, and up to $2\times$ larger than what the local approximation from \cref{vnqe} would predict for $T_e,n_e,$ and $\mathbf B$ profiles at 300 ps. This discrepancy decreases at higher field strengths due to magnetic field induced localization of the heat flow carrying electrons.

The degree of magnetic flux pile up in the hohlraum wall, however, is not so strongly affected by the increase in magnetic field strength because $\omega\tau \sim n_e^{-1}$. The magnitude of maximum field strength in the wall ranges from $2 < B_\text{y}/B_\text{y0} <3$ for $1 < B_\text{y0} < 100$ T. 

We have shown Vlasov-Fokker-Planck modeling of an external magnetic field of 1-100 T imposed upon a Omega-scale hohlraum. Magnetic flux pile-up causes an increase in magnetic field magnitude by a factor of 3 for a 7.5 T magnetic field. Additionally, the heat flow is responsible for magnetic field cavitation on a faster time-scale than that from the bulk flow of the plasma. Not only is the heat flow strongly non-local, it also has distinct signatures of inverse bremsstrahlung heating. The ability to preserve distribution function information through use of a kinetic code allows to model the heat flow accurately. Full Vlasov-Fokker-Planck-Maxwell treatment of the system enables accurate modeling of magnetic field dynamics. We have shown that the Nernst flow is the dominant mechanism for magnetic field transport and is responsible for the increase in field strength, up to 100 T for a initial 100 T field, in the wall as well as cavitation of the magnetic field towards the hohlraum axis. The magnetic field cavitation is mitigated at higher field strengths. Furthermore, the Nernst velocity is up to $2\times$ larger in VFPM than would be predicted by classical transport. 

These findings suggest that attempting the same calculation with a classical description of transport would result in significantly different $\vb B$ \& $T_e$ evolution. Accurate modeling of these quantities has implications for controlling levels of laser plasma interactions \cite{Montgomery2015,Regan2010} and hot electron propagation \cite{Strozzi2015} in the gas fill and understanding the hot spots on the dense wall that generate X-rays. The enhanced electron transport and $\vb B$ field physics presented here could affect details of X-ray drive if incorporated into full-scale radiation-hydrodynamics modeling (including reduced phenomenological laser-plasma interaction models) of indirect drive with externally applied B-field.

The authors would like to thank A. Hazi, J. Moody, D. Strozzi of LLNL and A. Sefkow of Sandia for useful discussions regarding related work at NIF and Omega. The modeling was performed using computational resources and services provided by Advanced Research Computing at the University of Michigan, Ann Arbor. This research was supported by the DOE through Grant No. DE SC0010621. 

\bibliography{../../../Papers/library.bib}

\end{document}